\documentclass[10pt, conference, compsocconf]{IEEEtran}
\usepackage{graphicx}

\usepackage{cite}

\ifCLASSINFOpdf
\else
\fi
\usepackage{array}

\usepackage{mdwmath}
\usepackage{mdwtab}

\usepackage{eqparbox}
\usepackage{fixltx2e}

\begin{document}
%
\title{Dynamic Learning of Pairwise and Three-way Entanglement}


\author{\IEEEauthorblockN{E.C. Behrman}
\IEEEauthorblockA{Department of Mathematics and Physics\\
Wichita State University\\
Wichita, KS 67260-0033,USA\\
elizabeth.behrman@wichita.edu}
\and
\IEEEauthorblockN{J.E. Steck}
\IEEEauthorblockA{Department of Aerospace Engineering\\
Wichita State University\\
Wichita, KS 67260-0044\\
james.steck@wichita.edu}
}


%


\maketitle

\begin{abstract}
In previous work, we have developed a dynamic learning paradigm for ``programming'' a general quantum computer.  A learning algorithm is used to find a set of parameters for a coupled qubit system such 
that the system at an initial time evolves to a state in which a given measurement 
results in the desired calculation value. This can be thought of as a quantum neural network (QNN).   Here, we apply 
our method to a system of three qubits, and demonstrate training the quantum computer to estimate both pairwise and three-way entanglement.
\end{abstract}


%
\IEEEpeerreviewmaketitle

\section{Introduction}


Detection and quantification of entanglement remains an important basic challenge in quantum information. Many different measures have been proposed\cite{bennett2,wootters,toth}. For a system of two qubits (quantum bits)  there does exist a general measure\cite{wootters}, but for larger systems\cite{greenberger} there are many problems. Most procedures require optimization\cite{vedral,tamaryan} and/or reconstruction of the density matrix\cite{park}, both of which can rapidly become impractical with increasing size of the system. Also, for systems larger than two qubits there exist inequivalent kinds of entanglement\cite{lohmayer}. The ``witness'' approach\cite{filip} is useful, but requires that the state of the system be ``close'', in some sense, to a given, known state.

Recently Behrman et al.\cite{behrman, behrmanqic} proposed using an approach based on adaptive computing: finding a set of parameters for the time evolution of a two-qubit system such that the qubit-qubit correlation function at the final time is mapped onto an entanglement witness. Remarkably, this witness did not require ``closeness'' to a given initial state. Here, we extend our work to a more complex three-qubit system. Our method succeeds in finding a single set of parameters such that all three qubit-qubit pairwise correlation functions give good witness for the pairwise entanglements for large classes of states. In addition, the three-point correlation function gives a good approximation to the three-way entanglement. No optimization procedure or prior state reconstruction is necessary.


\section{Dynamic Learning for the Coupled Three-qubit System: QNN}            
  
A quantum system, whether pure or mixed, evolves in time according to the Schrodinger equation:
\begin{equation}
\frac{d \rho}{dt} = \frac{1}{i \hbar}[H, \rho] 
\label{schr}
\end{equation}
where $\rho$ is the density matrix and $H$ is the Hamiltonian, whose formal solution\cite{peres} is 
\begin{equation}
\rho(t) = \exp (iLt) \rho (0).
\label{schrsoln}
\end{equation}
In practice, especially for time-varying Hamiltonians, the Schrodinger equation is usually integrated numerically; this is the approach we shall use here. We consider a three-qubit quantum system whose Hamiltonian is:
\begin{eqnarray}
H  & = &  K_{A} \sigma_{xA} + K_{B} \sigma_{xB} + K_{C} \sigma_{xC} \nonumber \\ 
   &   & \mbox{} + \varepsilon_{A} \sigma_{zA} + \varepsilon_{B} \sigma_{zB} + \varepsilon_{C} \sigma_{zC} \nonumber  \\
   &   & \mbox{} + \zeta_{AB} \sigma_{zA} \sigma_{zB} + \zeta_{AC} \sigma_{zA} \sigma_{zC} 
+ \zeta_{BC} \sigma_{zB} \sigma_{zC} 
\end{eqnarray}
where $\{ \sigma \}$ are the Pauli operators corresponding to each of the qubits, $\{K \}$ are the tunneling amplitudes, $\{ \varepsilon \}$  are the biases, and $\{ \zeta \}$, the qubit-qubit couplings. This represents three interacting qubits, labeled A, B, and C, respectively. We choose the so-called ``charge basis '', which for a system of three qubits is the set:
 $ |000\rangle$, $|001\rangle$, $|010\rangle$, $|011\rangle$,  $|100\rangle$, $|101\rangle$, $|110\rangle$, $|111\rangle$. The amplitude for each qubit to tunnel to its opposing state, i.e., switch between the 0 and 1 states, is its $K$ value; each qubit has an external bias represented by its $\varepsilon$ value; and each qubit is coupled to each of the other qubits, with a strength represented by the appropriate $\zeta$ value. We call $\varepsilon$ a ``bias'' because, if its value is positive, it will tend to force that qubit towards the $|0\rangle$ state (i.e., the energy for being in that state is lower), and if its value is negative, towards the $|1\rangle$ state. Similarly the sign of $\zeta$ lowers/raises the energy of the corresponding qubits to be aligned/antialigned. Note that, for example, $\sigma_{xA}$ acts only on qubit A. We can write that operator in the charge basis as an 8x8 matrix: $\sigma_{x} \otimes I  \otimes I$, where $\sigma_{x}$ is the familiar 2x2 matrix  $\left( \begin{array}{c c }  0   &   1 \\  1   &   0  \end{array} \right) $, $I$ the identity matrix $\left( \begin{array}{c c }  1   &   0 \\  0   &  1  \end{array} \right) $, and $ \otimes$ the outer product.  Similarly $\sigma_{zB}$ is the 8x8 matrix $ I  \otimes \sigma_{z} \otimes I $, where $\sigma_{z}$ is the matrix  $ \left(  \begin{array}{c c } 1   &   0 \\  0   &   -1 \end{array} \right) $. Any product term, e.g., $\sigma_{zA}\sigma_{zB}$, can also be written as a single matrix: $\sigma_{zAB} = \sigma_{z}  \otimes \sigma_{z} \otimes I$.

The parameters $\{ K,\varepsilon,\zeta \}$ control the time evolution of the system in the sense that, if one or more of them is changed, the way a given state will evolve in time will also change, because of Eq.~\ref{schr}. This is the basis for using our quantum system as a neural network. There is a mathematical isomorphism between Eq.~\ref{schrsoln} and the equation for information propagation in a neural network: $\phi_{output}= F_{W} \phi_{input }$, where $\phi_{output}$ is the output vector of the network, $\phi_{input}$ the input vector, and $F_{W}$ the network operator, which depends on the neuron connectivity weight matrix $W$. Here the role of the ``weights'' of the network is played by the parameters of the Hamiltonian, $\{ K,\varepsilon,\zeta  \}$, all of which can be adjusted experimentally as functions of time\cite{yamamoto, han}. By adjusting the parameters using a neural network type learning algorithm we can train the system to evolve in time to a set of particular final states at the final time $t_{f}$, in response to a corresponding set of given inputs. Because the time evolution is quantum mechanical (and, we assume, coherent), a quantum mechanical function like an entanglement witness can be mapped to an observable. Complete details, including a derivation of the quantum dynamic learning paradigm using backpropagation in time\cite{lecun,werbos}, are given in \cite{behrmanqic}.

The time evolution of the quantum system is calculated by integrating the Schrodinger equation in MATLAB 
Simulink \cite{matlab}. The ODE4 fixed step size solver was used with a integration step size of 0.05 ns. The system was initialized (prepared in) each input state, in turn, then allowed to evolve for 300 ns. All of the parameters were taken to be functions of time; this was done in the simulation by allowing them to change to a different constant value every 75 ns (i.e., four ``time chunks''.)  Discretization 
error for the numerical integration was checked by redoing the calculations with a timestep of a tenth the size; results were not affected.  Since the error needs to be back propagated through time, the integration has to be carried out from $t_{f}$ to 0.  To implement this in MATLAB Simulink, a change of variable is made by letting $t' = t_{f}-t$.

\setcounter{footnote}{0}
\renewcommand{\thefootnote}{\alph{footnote}}

\section{Pairwise Entanglement Witness}

In earlier work on a two-qubit system, we used a training set of only four pure states: the Bell, flat, a  correlated product, and a partially entangled state.  See Table I.  Though only trained on these four states, the net, when tested on states \underline{not} trained on, successfully reproduced a good approximation to the entanglement of formation for large classes of states, including mixed states and product states as well as fully and partially entangled states. We called this an ``experimental'' entanglement witness, because no prior knowledge of the state of the system was required: once the net was trained, simple time evolution under the Hamiltonian followed by a single measurement was sufficient to detect entanglement for \underline{any} input state. Indeed, the QNN did better than this: under most conditions measurement gave an approximate amount of entanglement, as well.

For a three-qubit system, we decided to build on this earlier work directly, that is, with an initial training set that attempted only to learn the three distinct amounts of \underline{pairwise} entanglement present. Thus, we started with a training set of twelve input-output pairs: three copies of the set of four we had successfully used earlier for two qubits, one copy for each pair of qubits. The input states are shown in Table II, and the desired (target) outputs, in Table III. For example, the Bell state for qubits B and C (for which we would train the entanglement output for the BC pair to be equal to one) could be any (normalized)  state of the form: $ [ a_{0}|0\rangle +a_{1} |1\rangle]  \otimes [|00\rangle + |11\rangle ] $; we chose to use the $Bell_{BC}$ state $ [ \frac{1}{\sqrt{2}}|0\rangle ]  \otimes [|00\rangle + |11\rangle]$. Table IV shows the parameters. The starting values for qubits A and B are those found in our earlier work\cite{behrmanqic}; these were also used as the initial values for qubit C. As before, we took as our output the square of the two-qubit correlation function at the final time. Thus the target values for the pairwise entanglement between qubits B and C were trained to the output function 
$ \langle O(t_{f}) \rangle = \langle \sigma_{zB} (t_{f}) \sigma_{zC}(t_{f}) \rangle^{2} = [ tr (\rho (t_{f}) \sigma_{zB} \sigma_{zC})]^{2}$, and similarly for the pairs AB and AC. 

\begin{table*}[!t]  
\caption{Training set and data for the two-qubit (AB) system\cite{behrmanqic}.  The relative amplitudes (for the ket states) are given 
without normalization for clarity.  The Bell state is maximally entangled, and $P$ is 
partially entangled. The middle two are product states 
(flat = $(|0 \rangle + |1 \rangle)_{A}(|0 \rangle + |1 \rangle)_{B}$ and 
$Cr = | 0 \rangle_{A}( | 0 \rangle + \gamma | 1 \rangle)_{B}$) and thus have zero entanglement. The classical correlation is computed as $\langle \sigma_{zA}(0) \sigma_{zB}(0) \rangle$.  $Cr$ is
classically correlated but not entangled.  The next three columns show the entanglement as calculated by the 
methods of Bennett\cite{bennett2}  and of Vedral \cite{vedral}(using the von Neumann metric and the Bures metric.)  
Both distance measures have been normalized to unity. The last column shows the trained values from our previous work\cite{behrmanqic}, for $\theta = 0$ and $\gamma = 0.5$,  after 2000 epochs; RMS error = $1.08 \times 10^{-5}$.}
\begin{tabular}{|l| c c c c |c| c c c | c  c |}  \hline
State  &\multicolumn{4}{|l|}{Relative amplitudes of} & Classical &\multicolumn{3}{|l|}{Theoretical entanglement} &\multicolumn{2}{|l|}{QNN}  \\ \cline{2-5} \cline{7-9} \cline{10-11}
{} & $|00\rangle$ & $|01\rangle$ &$|10\rangle$ &$|11\rangle$ & Correlation & Bennett & von Neumann & Bures & target & output 
 \\ \hline
Bell  & 1 & 0 & 0 & $e^{i \theta}$ & 1 & 1 & 1 & 1 & 1 & 0.99997\\
flat  & 1 & 1 & 1  & 1 & 0 & 0 & 0 & 0 & 0 & $2.01 \times 10^{-6}$\\
$Cr$ & 1 & $\gamma$ & 0  & 0 & $\frac{1-|\gamma|^{2}}{1+|\gamma|^{2}}$ & 0 & 0 & 0 & 0 & $2.61 \times 10^{-5}$ \\
$P$  & 0 & 1 & 1  & 1 & $-1/3$ & 0.55 & 0.32 & 0.46 & 0.44317 & 0.44317  \\ \hline 
\end{tabular} 
\end{table*}

\begin{table*}[!t]
\caption{Input matrix for training the three-qubit QNN entanglement witnesses. Each column is an input state, showing amounts of each of the basis states. The first twelve states are the set used for the \underline{pairwise} witness; all thirteen were used for the second training set.}
\begin{tabular}{|l| l l l l l l l l l l l l |l|} \hline
Input       & $Bell_{AB}$  &  $Bell_{AC}$  &$Bell_{BC}$  & $flat_{AB}$   & $flat_{AC}$ & $flat_{BC}$ & $Cr_{AB}$  & $Cr_{AC}$ & $Cr_{BC}$ & $P_{AB}$ & $P_{AC}$ & $P_{BC}$ & $GHZ_{-}$ \\
\hline
$ |000\rangle $    & $\sqrt{2}$ & $\sqrt{2}$   &  $\sqrt{2}$ &  0.5 & 0.5 & 0.5  &  0                                      &   0                                 &  0                                               &   $\sqrt{3}$           &  $\sqrt{3}$   &  $\sqrt{3}$  & $\sqrt{2}$ \\
$ |001\rangle $    &   0              &         0          &   0               &  0    & 0.5 &  0.5 &  0                                       &     0                                 &       0                                        &     0                        &  $\sqrt{3}$  & $\sqrt{3}$ & 0 \\
$ |010\rangle $    &   0              &         0          &   0               &  0.5 & 0    &  0.5 &  0                                       &     0                                    &   $\frac{0.5}{\sqrt{1.25}}$ &  $\sqrt{3}$             &   0                & $\sqrt{3}$ & 0 \\
$ |011\rangle $    &   0              &         0          & $\sqrt{2}$  &  0    & 0    &  0.5 &  0                                        &    0                                       &   $\frac{1}{\sqrt{1.25}}$  &   0                           &   0                &   0 & 0 \\ 
$ |100\rangle $    &   0              &         0          &      0            &  0.5 & 0.5 &  0   &  $\frac{0.5}{\sqrt{1.25}}$  & $\frac{0.5}{\sqrt{1.25}}$  &  0                                       &    $\sqrt{3}$          &  $\sqrt{3}$   &  0 & 0 \\
$ |101\rangle $    &   0              &   $\sqrt{2}$   &   0          &  0        & 0.5 &  0   &   0                                       &   $\frac{1}{\sqrt{1.25}}$     &     0                                     &      0                       &  0                  &  0 & 0 \\
$ |110\rangle $    &  $\sqrt{2}$ &        0             &   0          &  0.5    & 0    &  0   &  $\frac{1}{\sqrt{1.25}}$  &    0                                         &     0                                       &   0                         & 0                   &  0 & 0 \\ 
$ |111\rangle $    &   0              &         0             &   0          &  0       & 0    &  0   &   0                                     &   0                                          &    0                                         &      0                      &   0                 &  0    & $-\sqrt{2}$  \\ \hline
\end{tabular} 
\end{table*}

\begin{table*}[!t]
\caption{QNN entanglement target and output matrices for \underline{training}. Each column is an input state corresponding to the respective column in Table II; each row corresponds to a target function for the specified entanglement function: AB, $\langle \sigma_{zA} (t_{f}) \sigma_{zB}(t_{f}) \rangle^{2} $; AC, $\langle \sigma_{zA} (t_{f}) \sigma_{zC}(t_{f}) \rangle^{2}$; BC, $\langle \sigma_{zB} (t_{f}) \sigma_{zC}(t_{f}) \rangle^{2} $; and ABC, $\langle \sigma_{zA} (t_{f}) \sigma_{zB}(t_{f})\sigma_{zC}(t_{f}) \rangle^{2} $. The first three rows are the targets (desired values)for the first training set (pairwise entanglement only - the first twelve states shown in Table II); the second three, the trained values for that set; the next four, the targets for the second training set (all thirteen states shown in Table II); the last four, the trained values for that set. Nonzero target values are boldfaced for easy comparison. }  
\begin{tabular}{|l| l l l l l l l l l l l l l| l|} \hline
 &  & $Bell_{AB}$  &  $Bell_{AC}$  &$Bell_{BC}$  & $flat_{AB}$   & $flat_{AC}$ & $flat_{BC}$ & $Cr_{AB}$  & $Cr_{AC}$ & $Cr_{BC}$ & $P_{AB}$ & $P_{AC}$ & $P_{BC}$ & $GHZ_{-}$ \\ \hline
Targets: & AB & \bf{1}  & 0  & 0  & 0 & 0&  0 &  0 &  0 & 0 & \bf{0.44} & 0&  0    & - \\
Set 1 & AC & 0 & \bf{1} &  0 & 0 & 0 & 0 & 0 & 0 & 0 & 0 &  \bf{0.44} & 0  & - \\
& BC & 0 & 0 & \bf{1} & 0 & 0 & 0 & 0 & 0 & 0 & 0 & 0 & \bf{0.44} & - \\ \hline 
Trained & AB & \bf{}0.9943 & 0.0016 & 0.0001 & 0.0003 & 0.0018 & 0.0001 & 0.0007 & 0.0001 & 0.0002 & \bf{0.4399} & 0.0008 & 0.0003 & - \\
& AC & 0.0000 & \bf{0.9930} & 0.0001 & 0.0006 & 0.0001 & 0.0012 & 0.0003 & 0.0002 & 0.0018 & 0.0010 & \bf{0.4385} & 0.0016 & - \\
& BC & 0.0000 & 0.0017 & \bf{0.9945} & 0.0000 & 0.0009 & 0.0000 & 0.0013 & 0.0048 & 0.0002 & 0.0000 & 0.0000 & \bf{0.4392} & - \\ \hline  
Targets: & AB & \bf{1}  & 0  & 0  & 0 & 0&  0 &  0 &  0 & 0 & \bf{0.44} & 0&  0    & 0 \\
Set 2 & AC & 0 & \bf{1} &  0 & 0 & 0 & 0 & 0 & 0 & 0 & 0 &  \bf{0.44} & 0  & 0 \\
& BC & 0 & 0 & \bf{1} & 0 & 0 & 0 & 0 & 0 & 0 & 0 & 0 & \bf{0.44} & 0 \\ 
& ABC & 0 & 0 & 0     & 0 & 0 & 0 & 0 & 0 & 0 & 0 & 0 & 0 & \bf{1} \\ \hline 
Trained & AB & \bf{0.9911} & 0.0012 & 0.0000 & 0.0002 & 0.0023 & 0.0012 & 0.0010 & 0.0006 & 0.0005 & \bf{0.4384} & 0.0016 & 0.0014  & 0.0039\\
& AC & 0.0002 & \bf{0.9964} & 0.0000 & 0.0015 & 0.0001 & 0.0018 & 0.0000 & 0.0017 & 0.0001 & 0.0049 & \bf{0.4386} & 0.0021 & 0.0001 \\
& BC & 0.0002 & 0.0012 & \bf{0.9969} & 0.0002 & 0.0001 & 0.0000 & 0.0000 & 0.0009  & 0.0000 & 0.0026 & 0.0006 & \bf{0.4396} & 0.0066 \\
& ABC & 0.0012 & 0.0000 & 0.0028 & 0.0001 & 0.0014 & 0.0002 & 0.0000 & 0.0001  & 0.0003 & 0.0006 & 0.0024 & 0.0007 & \bf{0.9883} \\
\hline 
\end{tabular} 
\end{table*}

\begin{table*}[!t]
\caption{Parameters for entanglement, in MHz. Each column is a different parameter, as labelled. Each parameter is a function of time; the first column shows the timeslice number. The first four rows are the initial parameters, taken from our earlier work \cite{behrmanqic}; the next four, the parameters after training (5000 epochs; RMS error = $1.215 \times 10^{-3}$) to the first training set,  of twelve, for just pairwise entanglement  (the inputs shown in Table II, using the target values in Table III); the last four, after training to the second set, of thirteen (the twelve pairwise states plus one three-way.) Training for the second set was done for 5000 additional epochs; RMS error = $1.801\times 10^{-3}$.) }
\begin{tabular}{|l|l| c c c c c c c c c |} \hline 
& Timechunk & $K_{A}$ & $K_{B}$ &  $K_{C}$ &  $\varepsilon_{A}$ &  $\varepsilon_{B}$ & $\varepsilon_{C}$ & $\zeta_{AB}$ & $\zeta_{AC}$  & $\zeta_{BC}$ \\ \hline
Initial & 1 &  2.3576   &  2.3576 &  2.3576  &  0.10913   & 0.10913   & 0.10913     &  0.04503      &  0.04503      &  0.04503  \\
& 2 &  2.3576   &  2.3576 &  2.3576  & 0.03768  & 0.06377   & 0.06377     &  0.10117     &  0.10117     &  0.10117 \\
& 3 &  2.3577   &  2.3576 &  2.3576 & 0.08671    & 0.03880   & 0.03880     &  0.10771      &  0.10771      &  0.10771 \\ 
& 4 &  2.3461   &  2.3546 &  2.3546  & 0.07146   & 0.07239  & 0.07239     &  0.04422     &  0.04422     &  0.04422 \\ \hline
Set 1 & 1 & 2.0133 & 2.4955 & 2.4376 & -0.08676 & -0.44110 &  0.32887 &  -0.35415     & 0.06745 &  -0.08465 \\
& 2 & 2.2238 & 2.0328 & 2.4116 & -0.11451 & -1.08901   &  -0.39170 &   0.81660    & 0.22532 & -0.27369  \\
& 3 & 2.4349 & 2.2618 & 2.4203 & -0.05961 & -0.31644 &  -0.04549 &  -0.61235    & -0.33585 & 0.24552 \\
& 4 & 2.5836 & 2.3423 & 2.3853 &  0.03040 &  0.05885 & -0.02624 &   0.28887   & -0.07233 & 0.13128  \\ \hline

Set 2 & 1  & 2.1024  &  2.5880 & 2.3032 & 0.39188 &  0.55015 &  0.79315 &  -0.62404   & -0.26318    & -0.41928 \\
& 2  & 2.4013 &  2.1874 &  2.3051 & -0.04748 & -0.38236 & -0.37284 & 0.64560  &  0.21584   & -0.09610 \\
& 3  & 2.4969 &  2.2845 &  2.3047 &  -0.53340 & -0.48951 &  -0.34083 &  -0.97289 & -0.92255   & -0.74675 \\
& 4  & 2.5908 &  2.4842 &  2.4072 &  -0.31082 &  -0.10772  &  -0.16904 &  0.26100  &  -0.17536  & 0.01217  \\ \hline
\end{tabular}
\end{table*}

Training results are shown in Tables III and IV for the outputs and the trained parameters, respectively. RMS error per output for the set of 36 outputs (three for each member of the training set of 12), after 5000 passes through the training set, was $1.215 \times 10^{-3}$. To determine whether the net has actually learned, we then tested on a large number of states not included in the training set, including fully entangled states, partially entangled states, product (unentangled) states, and also mixed states. A representative sample is shown in Table V. RMS error per output for the set of 39 outputs was $3.129 \times 10^{-2}$. While errors are significantly larger, it is also clear from the output matrix that there is definite separation, in two senses: first, that it is easy to see where the pairwise entanglement is (e.g., to distinguish between a state with AB entanglement and one with BC entanglement); and, second, that it is easy to tell the difference among unentangled, partially entangled, and fully entangled states. In other words, while this is not, strictly speaking, an entanglement measure, it is quite a good witness.

\begin{table*}[!t]
\caption{Representative QNN pairwise entanglement \underline{testing}: target and output matrices. No additional training was done: merely, the outputs were evaluated using the trained parameters for time evolution in Table IV, for a new set of input states. Each column is an input state. The first three are EPR states of the form $\frac{1}{\sqrt{2}}(|01\rangle \pm |10 \rangle)$ for qubits AB, AC, and BC, times a superposition state for the third qubit: $\frac{1}{\sqrt{2}} (|0\rangle + |1 \rangle)$; the next three, product (unentangled) states (the fourth is  the state $\frac{1}{\sqrt{8}}(|000\rangle + |001\rangle + |010\rangle + |011\rangle + |100\rangle + |101\rangle + |110\rangle + |111\rangle)$; the fifth, the state $|000\rangle$; the sixth, the product state $0.64(0.8|0\rangle+|1\rangle)|1\rangle(|0\rangle+0.7|1\rangle)$); the next three, P' states of the form $|00\rangle \pm |11 \rangle$ for qubits AB, AC, and BC, times a superposition state for the third qubit: $\frac{1}{\sqrt{2}} (|0\rangle + |1 \rangle)$; the next three, Bell states times a superposition state; and the last, a (completely) \underline{mixed} state: $M_{\pm} = 0.5(|000\rangle\langle 000| \pm 
|111\rangle \langle 111|)$.  Each row corresponds to a value for the specified entanglement function: $\langle \sigma_{zA} (t_{f}) \sigma_{zB}(t_{f}) \rangle^{2} $,  $\langle \sigma_{zA} (t_{f}) \sigma_{zC}(t_{f}) \rangle^{2}$, and $\langle \sigma_{zB} (t_{f}) \sigma_{zC}(t_{f}) \rangle^{2} $, respectively. The first three rows are the targets (desired values); the last three, the calculated outputs, using the parameters in Table III for the first (pairwise) training set. Upper vs lower signs made essentially no difference to the calculated values. Nonzero target values are boldfaced for easy comparison.}  
\begin{tabular}{|l|l| l l l l l l l l l l l l l|} \hline

& & $E_{AB}$ & $E_{AC}$ &$E_{BC}$ & $F_{1}$ & $F_{2}$& $F_{3}$ & $P'_{AB}$ & $P'_{AC}$ & $P'_{BC}$ & $Bl_{AB}$ & $Bl_{AC}$ & $Bl_{BC}$ & $M$ \\
\hline
Targets & AB & \bf{1} & 0 & 0 & 0 & 0 & 0 & \bf{0.44} & 0&  0 & \bf{1} & 0 & 0 & 0  \\
& AC & 0 & \bf{1} & 0 & 0 & 0 & 0 & 0 & \bf{0.44} & 0 & 0 & \bf{1} & 0 & 0  \\
& BC & 0 & 0 & \bf{1} & 0 & 0 & 0 & 0 & 0 & \bf{0.44} & 0 & 0 & \bf{1} & 0 \\ \hline
Outputs & AB & \bf{0.9365} & 0.0000 & 0.0004 & 0.0019 & 0.0006 & 0.0051 & \bf{0.4798} & 0.0022 &  0.0001 & \bf{0.9661} & 0.0005 & 0.0004 & 0.0002 \\
& AC & 0.0000 & \bf{0.7991} & 0.0023 & 0.0010 & 0.0002 & 0.0006 & 0.0003 & \bf{0.4034} &  0.0017 & 0.0028 & \bf{0.9227} & 0.0009 & 0.0004 \\
& BC & 0.0005 & 0.0001 & \bf{0.9419} & 0.0080 & 0.0013 & 0.0081 & 0.0011 & 0.0022 &  \bf{0.3276} & 0.0026 & 0.0002 & \bf{0.7692} &  0.0010  \\ \hline 
\end{tabular} 
\end{table*}
\section{Three-way entanglement} 

Clearly the net has been trained successfully to calculate an approximation to pairwise entanglement. But can it distinguish between pairwise and three-way entanglement? To find out, we then tested the pairwise-trained net on the GHZ states $\frac{1}{\sqrt{2}}(|000 \rangle \pm |111 \rangle)$. All the pairwise entanglements tested as very small: $\langle \sigma_{zA} (t_{f}) \sigma_{zB}(t_{f}) \rangle^{2}=  0.0031$; $\langle \sigma_{zA} (t_{f}) \sigma_{zC}(t_{f}) \rangle^{2}= 0.0032$; and  $\langle \sigma_{zB} (t_{f}) \sigma_{zC}(t_{f}) \rangle^{2}= 0.0015$, while the three-point correlation function at the final time was significantly  \underline{non}zero: $ \langle \sigma_{zA} (t_{f}) \sigma_{zB} (t_{f}) \sigma_{zC}(t_{f}) \rangle^{2} = 0.57$. This seemed to indicate that the QNN was able to distinguish between pairwise entanglement, as exhibited in the Bell and EPR states, and the inequivalent, three-way entanglement of a GHZ state. Perhaps the net could be trained to find yet another output.  Buoyed by hope and encouraged by these results, we then proceeded to add one additional output for the net to train, the three-point correlation function at the final time, $\langle \sigma_{zA} (t_{f}) \sigma_{zB} (t_{f}) \sigma_{zC}(t_{f}) \rangle^{2}$, and one additional training pair to the input matrix: $|GHZ_{-}\rangle = \frac{1}{\sqrt{2}}(|000 \rangle - |111 \rangle)$, for which we set the target outputs as 0, 0, 0, and 1 (for each of the two point functions and the three point function, respectively: see Table III.) The net was then retrained, beginning from the values of the parameters found in the previous section, with the new training set of thirteen. Results are shown in Table III, and the new values for the parameters are shown in Table IV. Testing on product states and on pairwise-entangled states was repeated; though errors were slightly larger, results were substantially similar. We also tested $|GHZ_{+}\rangle = \frac{1}{\sqrt{2}}(|000 \rangle + |111 \rangle)$; the outputs for the pairwise entanglements and for the three-way entanglement were indistinguishable from the outputs for the trained state  $|GHZ_{-}\rangle$. The net now seems to be able to recognize both pairwise \underline{and} three-way entanglement. 

If we had an experimental realization of the three-qubit system, we could use these results to find (approximately) a pairwise or three-way entanglement for any (initial) state of the system, by setting the various parameters to the determined values, allowing the system to evolve for the set time, then measuring the appropriate correlation function. Of course, any real experimental system will have nonzero sources of error, like noise and decoherence, which will mean the results of the simulation will not be exactly correct. One of the great advantages of the neural network approach is its ability to deal easily with complications like these: with ``online'' training, the QNN can take these into account in the training phase, adjusting the values of the parameters accordingly.  

Within the theoretical simulation, it is a simple matter to calculate the entanglement of any state we wish.  Some interesting results are shown in the two figures. In Figure 1, we plot the entanglement, as calculated by the QNN, for the state $\alpha |000\rangle + \beta|001 \rangle + |010\rangle + |100\rangle$ , as a function of both $\alpha$ and $\beta$. The graph shows two kinds of pairwise entanglement, AB (red) and AC (blue), and three-way entanglement (green). The three-way entanglement (green) is zero on the scale of the graph for all states plotted; the other pairwise entanglement, BC, lies right on top of the AC (blue) entanglement on this scale, and so is omitted for clarity. Along the $\alpha$ axis, when $\beta = 1$ (left hand edge), the pairwise entanglements AB (red) and AC (blue) are equal (i.e., the surfaces meet); this must be true, of course, by symmetry.  When $\alpha = 0$ and $\beta = 0$ (front right corner), this is the EPR state in qubits A and B, outerproducted with $|0\rangle_{C}$; its AB pairwise entanglement (red) is maximal while other pairwise entanglements (blue), as well as the three-way entanglement (green), are zero. When $\alpha = 0$ and $\beta = 1$ (front corner), this is the so-called ``W''\cite{tamaryan} state (the three-qubit generalization of the EPR state), in which entanglement is shared among all three qubits, but in a dissimilar manner to that in the GHZ state\cite{GHZvsW}.  The net calculates that the W state has all three kinds of pairwise entanglement, in (approximately) equal amounts: $\langle \sigma_{zA} (t_{f}) \sigma_{zB}(t_{f}) \rangle^{2}=  0.3899$; $\langle \sigma_{zA} (t_{f}) \sigma_{zC}(t_{f}) \rangle^{2}= 0.4129$; and  $\langle \sigma_{zB} (t_{f}) \sigma_{zC}(t_{f}) \rangle^{2}=  0.4042$, while the three-way (GHZ) entanglement is calculated to be essentially zero:  $\langle \sigma_{zA} (t_{f}) \sigma_{zB}(t_{f})\sigma_{zC} (t_{f}) \rangle^{2} =1.7628 \times 10^{-3}$ . In the other direction, at $\alpha = 1$ and $\beta = 0$ (back corner), this is the two-qubit partially entangled ``P'' state for the AB pair, and the net correctly calculates that the AB entanglement (red) is about 0.4 while the other pairwise entanglements (blue) are approximately zero (see Table III for precise numbers.)

\begin{figure} [!t]
\includegraphics[height=3in]{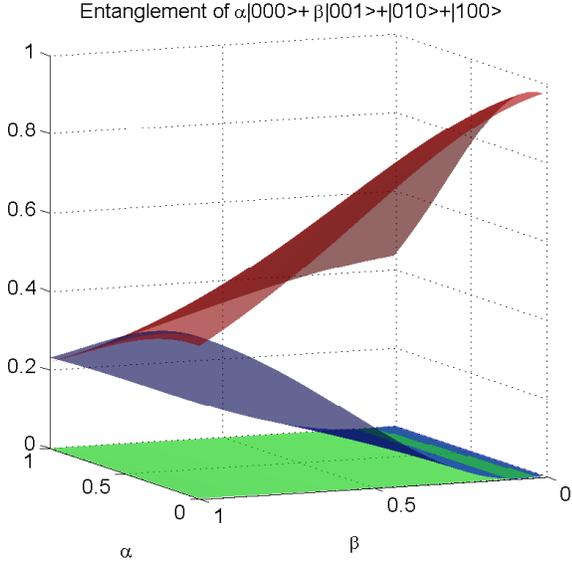} 
\caption{\label{WandP} Entanglement of the pure state 
$\frac{1}{\sqrt{2+\alpha^{2}+\beta^{2}}}(\alpha|001\rangle + \beta|000\rangle + |010\rangle + |100\rangle)$ , 
as a function of $\alpha$ and $\beta$, as calculated by the QNN, using the trained parameters listed in Table IV. 
Each color represents a different entanglement measure: red, the pairwise entanglement between qubits A and B; 
blue, that between A and C; and green, three-way entanglement among A, B, and C.  }
\end{figure}

Figure 2 shows three-way (GHZ) entanglement (green) and pairwise AB entanglement (red), for the state $\alpha|110\rangle + \beta|111\rangle + |000\rangle$, as a function of both $\alpha$ and $\beta$. For $\alpha = 0$ and $\beta = 0$, this is the product $|000\rangle$ state, for which pairwise (red) and three-way entanglement (green) are both zero; for $\alpha = 1$ and $\beta = 0$, this is the $Bell_{AB}$ state (outer producted with $|0\rangle_{C}$), $(|11\rangle  + |00\rangle)\otimes |0\rangle$, which has maximal AB entanglement (red) but zero three-way entanglement (green) ; when $\alpha = 0$ and $\beta = 1$, this is the GHZ state $ |111\rangle + |000\rangle$, which has zero AB entanglement (red) and maximal three-way entanglement (green). The two surfaces can be seen to cross each other along the black line which is projected onto the $\alpha\beta$ plane; on the scale of these calculations it is indistinguishable from $\alpha=\beta$.

\begin{figure} [!t]
\includegraphics[height=3in]{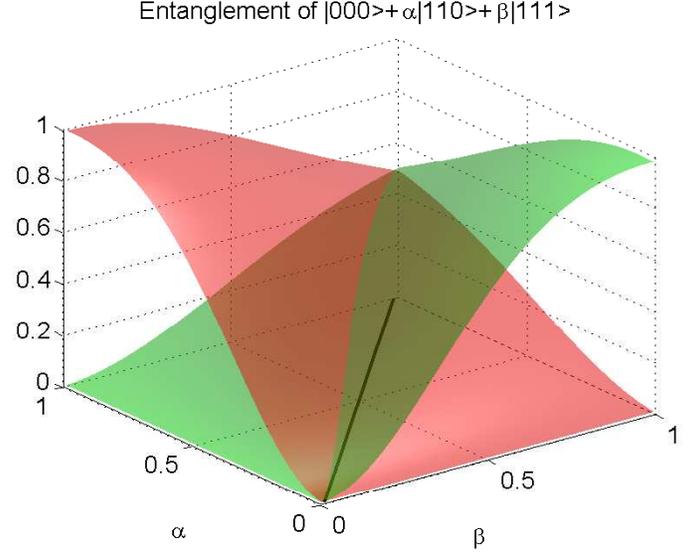} 
\caption{\label{GHZandBell} Entanglement of the pure state 
$\frac{1}{\sqrt{1+\alpha^{2}+\beta^{2}}}(\alpha |110\rangle + \beta|111\rangle + |000\rangle)$, 
as a function of $\alpha$ and $\beta$, as calculated by the QNN, using the trained parameters listed in Table IV. The three-point correlation function, which represents three-way entanglement, is in green; pairwise AB entanglement, in red. The line along which the AB entanglement is equal to the ABC entanglement is in black, projected onto the $\alpha\beta$ plane.}
\end{figure}

\section{Conclusion}
We have shown that a quantum computer of three qubits can be ``trained'' to compute, approximately, its own degree of entanglement. Pairwise and three-way entanglement can each be determined. No prior state reconstruction or tedious optimization procedure is necessary, nor is ``closeness'' to any particular state; rather, \underline{any} unknown state's entanglement can be estimated, whether pure or mixed.  

Further work that extends these results to the four-qubit system, with comparisons to the three-tangle\cite{coffman} and other measurements\cite{kimble}, is in progress.


\section*{Acknowledgment}

This work was supported in part by the National Science Foundation under Grant No. NSF PHY05-51164, through the the KITP Scholars program (ECB). We thank J.F. Behrman for extremely helpful discussions.




\begin{thebibliography}{19}
\bibitem{GHZvsW}
This is easy to see as follows.  Suppose the state of the system is  $|001 \rangle +  |010 \rangle +  |100 \rangle$, and the A qubit is measured. Then if the result of the measurement is 0,  qubits B and C are still fully entangled - in fact, they are in the EPR state, $|01 \rangle +  |10 \rangle $.  (If the result were 1, then the state of the system becomes $|00 \rangle$, which is of course not entangled.) On the other hand, suppose the state of the system is 
$ |000 \rangle +  |111 \rangle$, and again we measure A. It is obvious that no matter the result, the entanglement is completely destroyed, with the state of the system becoming either $ |00 \rangle $ or  $ |11 \rangle$, depending on the result of the measurement.
\end{thebibliography}
%

\end{document}